\documentstyle[sprocl,psfig,epsf]{article}

\def\Journal#1#2#3#4{{#1} {\bf #2}, #3 (#4)}

\def\NPB{{\em Nucl. Phys.} B}
\def\NPA{{\em Nucl. Phys.} A}
\def\PLB{{\em Phys. Lett.}  B}
\def\PRL{\em Phys. Rev. Lett.}

\def\PRC{{\em Phys. Rev.} C}
\def\EPC{{\em Eur. Phys. J.} C}
\def\ZPC{{\em Z. Phys.} C}

\def\be{\begin{equation}}
\def\ee{\end{equation}}
\def\bea{\begin{eqnarray}}
\def\eea{\end{eqnarray}}

\begin{document}


\smallskip

\title{PION--NUCLEON SCATTERING AND ISOSPIN VIOLATION}

\author{Ulf-G. Mei{\ss}ner}

\address{Forschungszentrum J\"ulich, Institut f\"ur Kernphysik (Theorie)\\ 
D-52425 J\"ulich, Germany\\E-mail: Ulf-G.Meissner@fz-juelich.de}

\maketitle\abstracts{I discuss low--energy pion--nucleon scattering
in the framework of chiral perturbation theory. I argue that using this 
theoretical method one is able to match the in some cases impressive
experimental accuracy (for the low partial waves). I then show how
strong and electromagnetic isospin violation can be treated simultaneously.
Some first results for neutral pion scattering and the $\sigma$--term are
given.}

\section{Introduction}

Arguably the biggest success of chiral symmetry encoded in the current
algebra of the sixties was the prediction for the isovector and isoscalar
S--wave pion--nucleon scattering lenghts,
$\tilde{a}^- = {M_\pi^2}/{8\pi F_\pi^2}  = 8.96$, $\tilde{a}^+ = 0$,\cite{wein}
in units of $10^{-2}/M_\pi$ (which will be used throughout). Here, $M_\pi$
and $F_\pi$ are the charged pion mass and the pion decay constant, in order.
The agreement with the then accepted empirical values of $\tilde{a}^- = 9\pm
1$, $\tilde{a}^+ = 0 \pm 1$ is quite 
impressive. However, while there have been many attempts to calculate the
corrections to these current algebra predictions by invoking hard pion techniques,
unitarization, super--propagator methods and so on, no definite conclusions on 
the size and even sign of these
corrections could be drawn. In fact, these lowest order predictions can now be
found in many textbooks on particle physics and quantum field theory (QFT) as an
example how symmetries can be used to make model--independent predictions without
a full understanding of the non--perturbative dynamics. For the case at hand,
this symmetry is the spontaneously broken chiral symmetry of the
strong interactions.
The not yet solved underlying QFT is, of course, {\em QCD}.
With the advent of chiral perturbation theory (CHPT), it has become possible
to readdress the question concerning the corrections to the abovementioned
predictions. Also,
impressive progress has been made in the measurements of the level shifts and
broadening in pionic hydrogen and deuterium, leading to much improved values
for the zero energy scattering amplitude, alas the S--wave scattering lengths.
In this talk, however, I will mostly be concerned with the theoretical developments
without going into technical details, rather I will address a collection of
objections, questions and misconceptions frequently encountered.

\section{Pion--nucleon scattering to one loop in CHPT}

CHPT is a systematic low--energy expansion for any strong interaction
process. This goes along  with an expansion in pion loops
supplemented by all local contact interactions allowed by chiral and
other symmetries. The latter terms are accompanied by the so--called
low--energy constants (LECs). The general strategy is to fit these by some data and
then move on to make predictions. It is often claimed that the
increasing number of these  LECs makes CHPT useless beyond some low
order. For example, the number of LECs of the most general two flavor
effective chiral pion--nucleon Lagrangian coupled to external fields
(in the heavy fermion formulation) 
at dimension two, three and four is 7$\,$\cite{bkkm}, 31$\,$\cite{em,fms} 
and $\sim 160\,$\cite{gm}, in
order.\footnote{For dimension four, the number given refers to the
LECs needed for renormalization only.} 
A variety of calculations performed
so far have shown that with some exceptions one has to go to fourth
order to achieve a good theoretical precision. So let us see how many
terms can contribute to $\pi N$ scattering. The pertinent T--matrix
is most conveniently described in terms of the standard invariant
amplitudes
\be
T^\pm_{\pi N} = A^\pm + q^2 \, B^\pm ~,
\ee
in a highly symbolic notation, with $q^2$ the squared
momentum transfer. The invariant amplitudes are functions
of two variables, which one can choose to be $\nu$ and $t$. Note
also that $\nu$ and $t$ count as ${\cal O}(p)$ and ${\cal O}(p^2)$,
respectively, with $p$ denoting a genuine small momentum. The most
general polynom for the four amplitudes $A^\pm, B^\pm$
to fourth order commensurate with crossing and the
other symmetries thus takes the form 
\bea
A^+_{\rm pol} &=& a_1 + a_2 t + a_3 \nu^2 + a_4 t^2 + a_5 t \nu^2 +
a_6 \nu^4~, \,\,\, 
A^-_{\rm pol} = \nu \, (b_1 + b_2 t + b_3 \nu^2 )~, \nonumber \\
B^+_{\rm pol} &=& c_1 \nu~, \,\,\, B^-_{\rm pol} = d_1 + d_2 t + d_3 \nu^2~,
\eea
so that in total we have 14 LECs since at third order there is one
more related to the Goldberger--Treiman discrepancy, i.e. a local
term with a LEC which allows
to express the axial--vector coupling $g_A$ in terms of the
pion--nucleon coupling $g_{\pi N}$. While the former appears naturally
in the effective Lagrangian, the latter is more suitable to discuss 
$\pi N$ scattering. So if one calculates to orders $p^2$, $p^3$ and
$p^4$, one has to pin down 5, 9 and 14 LECs, respectively. This
pattern is quite different from the total number of terms allowed at
the various orders, but it is a general rule that simple processes
do not involve any exorbitant number of LECs. In case of $\pi N \to
\pi N$, given the large body of data, one clearly has predictive power.
Some additional general remarks are in order here. First, one can
only expect to get precise predictions from CHPT for the low partial
waves in the threshold region. This is simply related to the fact that
because of angular momentum barrier factors, the higher the partial
wave, the higher the power in $p$ is where it starts to have a
nonvanishing contribution. Also, if one attempts to describe the
partial waves with pronounced resonances, like e.g. the $P_{33}$ one
dominated by the $\Delta (1232)$, one either has to stay well below the resonance
energy or to include these resonances in a systematic fashion. In what 
follows, I will exclusively discuss the first option.

\begin{figure}[t]
\psfig{figure=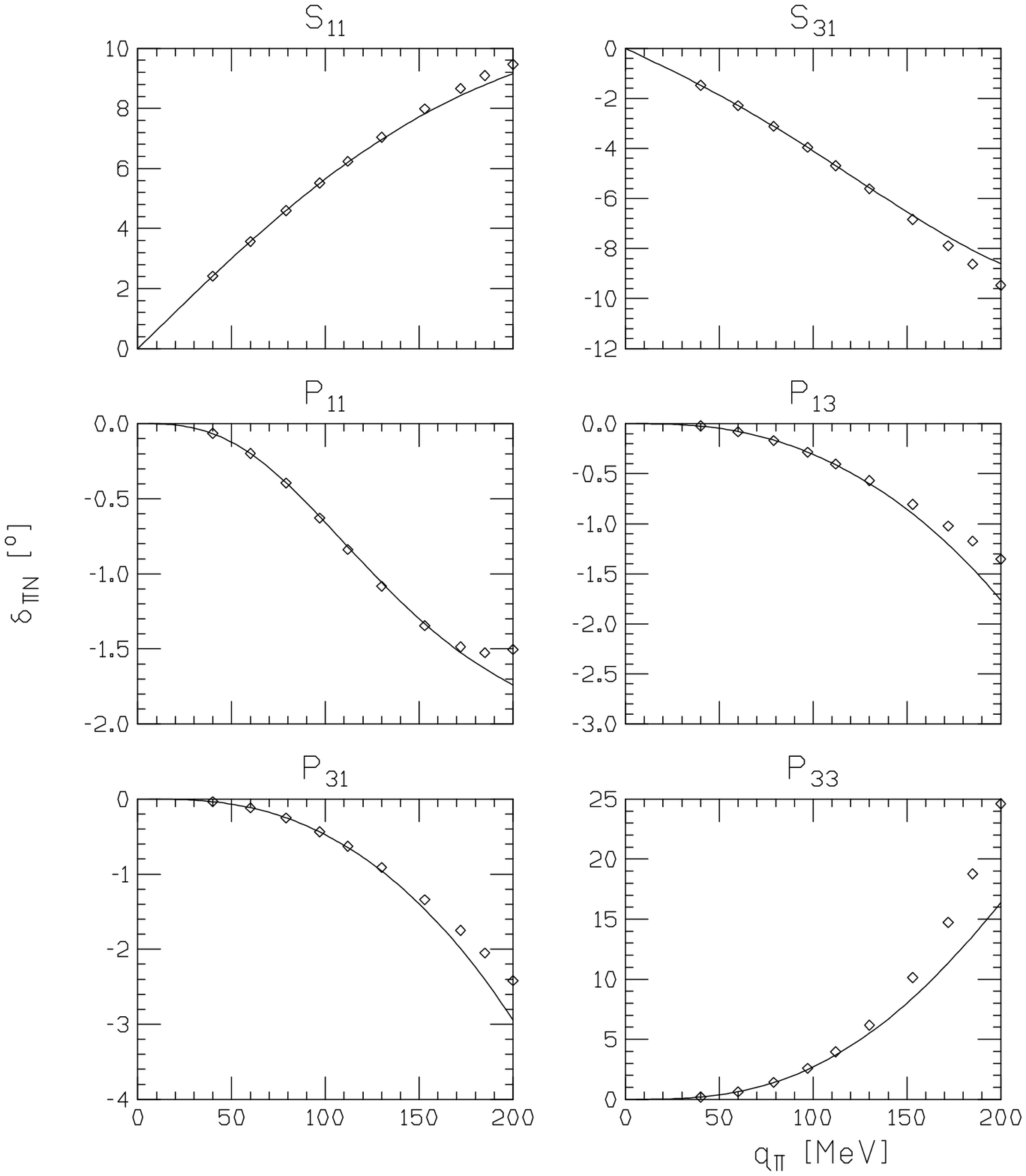,height=2.6in}
\vskip -2.6in
\hfill
\psfig{figure=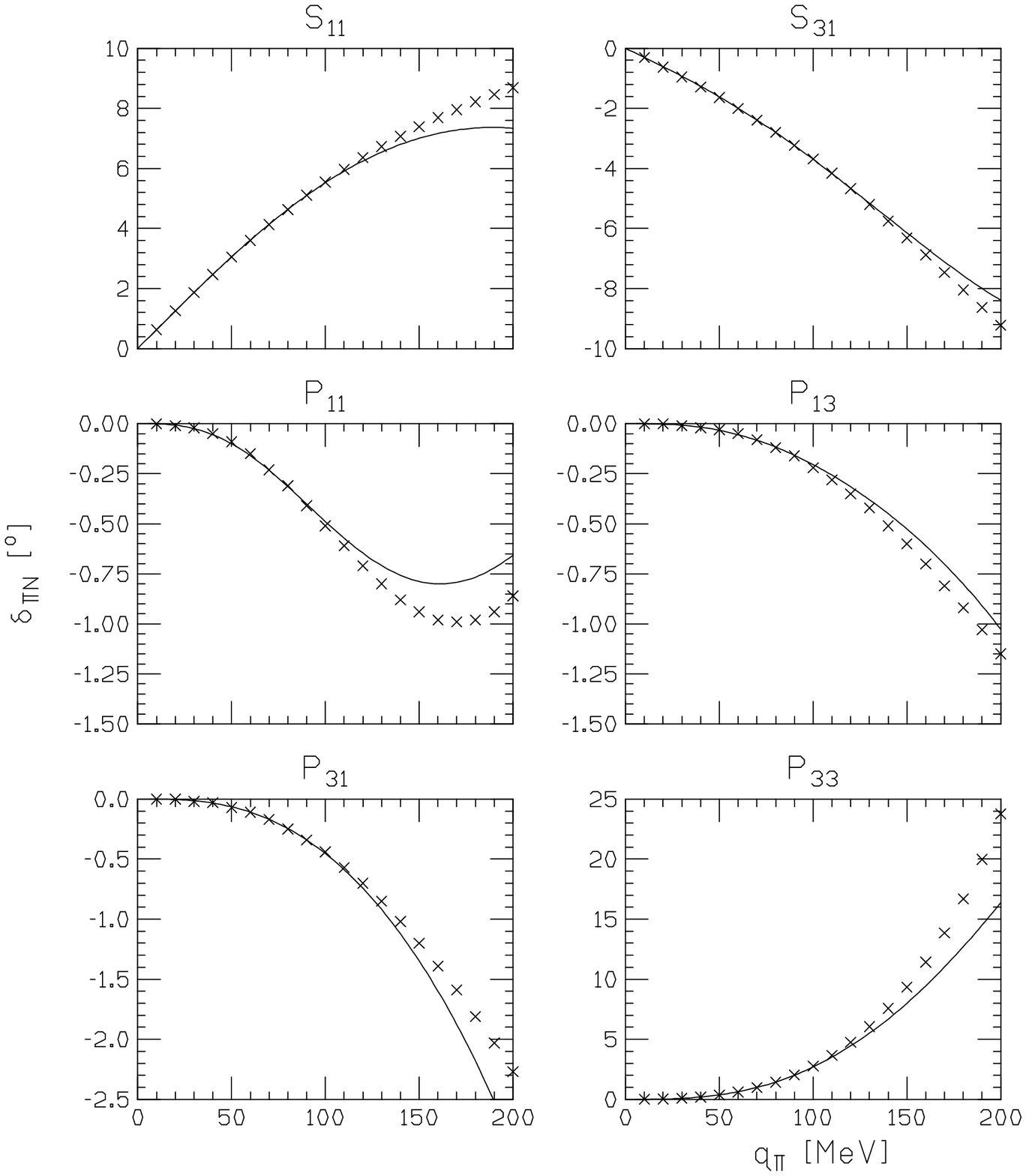,height=2.6in}
\caption{Fits and predictions for the S-- and P--wave phase shifts based
on the KA85 (left panel) and VPI SP98 (right panel) solutions. $q_\pi$ 
is the pion lab momentum. Fitted in each
partial wave are the data between 40 and 97~MeV (KA85) and 60 to
100~MeV (SP98). For lower and higher momenta, the phases are predicted.
\label{fig:phase}}
\end{figure}
Consider now CHPT in the isospin limit ($m_u = m_d$) and in the
absence of electromagnetism ($e = 0$), i.e. in the isospin symmetric world.
The first calculation of the chiral corrections to the S--wave scattering
lengths goes back to ref.\cite{bkmplb} and was sharpened in
ref.\cite{bkmprc}. Further aspects, including the one loop
contributions to ${\cal O}(p^3)$, were discussed in ref.\cite{bkmnpa}.
The complete $\pi N$ amplitude to third order was first discussed in 
ref.\cite{moj} (note that in relativistic nucleon CHPT, 
this amplitude was already given in ref.\cite{gss}).
The appearing nine LECs were fitted to the Karlsruhe--Helsinki threshold
parameters, the pion--nucleon $\sigma$--term and the
Goldberger--Treiman discrepancy. However, since there is some debate
about the actual value of the $\sigma$--term and also some of the
threshold parameters are not that precisely known, an alternative way
was followed in ref.\cite{fms}. There, the two S--  and six P--waves
from three different partial wave solutions were fitted for pion lab momenta
between 40 and 100~MeV, i.e. in the region where data exist, cf. Fig.1.
In particular, this allows to predict the partial waves at lower and
at higher energies. Using these three different inputs (which are
still under hot debate between the various
protagonists\footnote{Clearly, the dispersion theoretical analysis 
employed by 
the Karlsruhe--Helsinki group is the best method. One would therefore
like to see such a calculation repeated using also the more modern
data and an improved treatment of the electromagnetic corrections.}),  
one can get a handle on the theoretical uncertainty within
the order one calculates. The predictions for the
S--wave scattering lengths obtained in ref.\cite{fms} are (notice that
a kinematical factor $(1+\mu)^{-1}$, with $\mu$ the ratio of the pion
to the nucleon mass, has been included in the definition of $a^\pm$)
\be
8.3 \le a^- \le 9.3\,\, , \quad -1.0  \le a^+ \le 0.6\,\, ,
\ee
as shown by the box labelled ``CHPT'' in Fig.2. The theoretical
uncertainty reflects entirely the variation within the order
calculated, but does not account for possible effects due to higher
orders. Also shown are the
current algebra (CA) prediction and the empirical bands from the
measurements of the level shift in pionic hydrogen\cite{sigg} (the
band labelled ``H'') and pionic deuterium\cite{chat} (the band denoted
``D''). The width of the 1S level in pionic hydrogen, which gives another
bound on $a^-$, has also been measured but the analysis is not yet
finished due to some complications related to Doppler broadening effects.
\begin{center}
\begin{figure}[hbt]
\centerline{
\psfig{figure=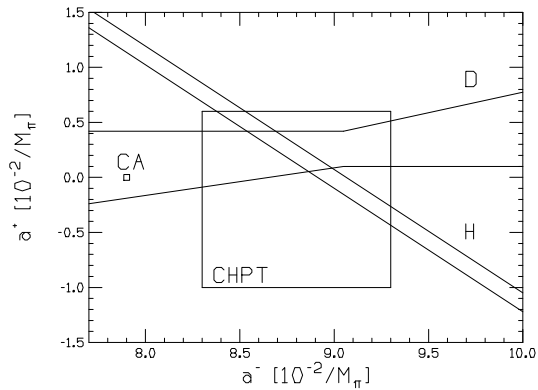,height=2.0in}
}
\caption{S--wave scattering lengths. Shown are the results obtained
from the level shift in pionic hydrogen (H--band) and pionic deuterium 
(D--band) compared to the
current algebra (CA)  and order $p^3$ CHPT prediction.
\label{fig:apm}}
\end{figure}
\end{center}
The present bound from the width (not shown in Fig.2) 
is $a^- = 8.8\pm 0.4$, consistent with the CHPT prediction.
The most important observation here is that to move from the CA point to the
empirical region requires pion loops, i.e. the shift from the CA point to
the CHPT box is largely a loop effect, as first noted in ref.\cite{bkmprc}.
Another constraint comes from charged pion photoproduction. The
threshold value of the electric dipole amplitude $E_{0+} (\gamma n \to
\pi^- p)$  is directly proportional to $a^-$. Taking the dispersion
theoretical result of ref.\cite{hdt} for $E_{0+}$ and the measured
Panofsky ratio, $P= \sigma (\pi^- p \to \pi^0 n)/ \sigma (\pi^- p \to
\gamma n) = 1.543 \pm 0.008$,  this leads to $a^- = 8.5 \pm
0.2$. There has also been a direct measurement of the inverse process
at TRIUMF. Its final analysis is eagerly awaited for. 

Concerning the P--waves, the situation is less satisfactory, which to
some extent is related to the $\Delta$ resonance being integrated out
in this approach and thus being hidden in the LECs of certain
operators. Still, when we are mostly concerned with the physics
encoded in the S--wave scattering lengths, this apparent deficiency
vanishes since in an approach with explicit $\Delta$'s, one would
have to worry about its mass and the $\Delta N \pi$ coupling
constant. The latter is
frequently derived from the width $\Gamma (\Delta \to N\pi)$, but a dispersive
sum rule for the $P_{33}$ partial wave, which allows to separate the
resonant and non--resonant contributions, leads to a by 40\% reduced
value.\cite{LB}

Of course, these calculations need to be improved. First, the convergence
of the chiral expansion is not very rapid to third order, so one definitively
has to calculate to order $p^4$. Second, 
the chiral expansion consists of even and odd powers in small momentum. 
It was argued in ref.\cite{juerg} that in fact the even and the odd series
do not talk to each other, which makes a one--loop ${\cal O}(p^4)$ calculation
even more mandatory, since it gives the first correction to the dimension
two tree graphs. The fourth order calculation is in progess.\cite{nadia} 

\section{Isospin violation in pion--nucleon scattering}

Let me start with some general remarks. Since a large body of elastic
scattering and charge exchange data exists, one has the possibility of
deducing bounds on isospin violation from simple triangle relations,
which link e.g. the processes $\pi^\pm p \to \pi^\pm p$ and $\pi^- p
\to \pi^0 n$. Care has, however, to be taken since there are two
sources leading to isospin violation. One is the ``trivial'' fact that
electromagnetism does not conserve I--spin, since it couples to the charge.
The other one is a strong effect, linked to the difference of the light
quark masses, $m_d -m_u$. This is essentially the quantity one is
after. In terms of the symmetry breaking part of the QCD Hamiltonian,
i.e. the quark mass term, we have
\be
{\cal H}_{\rm QCD}^{\rm sb} = m_u \bar{u}u + m_d \bar{d}d
= \frac{1}{2}(m_u + m_d) ( \bar{u}u +  \bar{d}d ) +
\frac{1}{2}(m_u - m_d) ( \bar{u}u -  \bar{d}d )~,
\ee
so that the strong I--spin violation is entirely due to the isovector term.
This observation lead Weinberg\cite{mass} to address the question of I-spin
violation in the pion and the pion--nucleon sector, with the remarkable
conclusion that in neutral pion scattering off nucleons one should expect
gross violations of this symmetry, as large as 30\%. Only recently
an experimental proposal to measure the $\pi^0 p$ scattering length in
neutral pion photoproduction off protons below the secondary $\pi^+ n$
threshold has been presented and we are still far away from a determination
of this elusive quantity.\cite{aron}
At present, there exist two phenomenological analysis~\cite{gibbs,mats}
which indicate isospin breaking as large as 7\% in the S--waves (and smaller
in the P--waves). Both of these analysis employ approaches for the strong
interactions, which allow well to fit the existing data but can not  easily
be extended to the threshold or into the unphysical region. What is, however,
most disturbing is that the electromagnetic corrections have been calculated
using some prescriptions not necessarily consistent with the strong interaction
models used. One might therefore entertain the possibility that some of the
observed I--spin violation is caused by the mismatch between the treatment
of the em and strong contributions. Even if that is not the case, both models
do not offer any insight into the origin of the strong isospin violation, but
rather parametrize them. In CHPT, these principle problems can be circumvented
by constructing the most general effective Lagrangian with pions, nucleons
and virtual photons,\cite{ms}
\be
{\cal L}_{\pi N} =  {\cal L}_{\pi N}^{(1)} +  {\cal L}_{\pi N,{\rm str}}^{(2)} +
{\cal L}_{\pi N,{\rm em}}^{(2)} + {\cal L}_{\pi N,{\rm str}}^{(3)} +
{\cal L}_{\pi N,{\rm em}}^{(3)} + \ldots~.
\ee
where the superscript gives the chiral dimension. It is important to note that
the electric charge counts as a small momentum, based on the observation that
$e^2/4\pi \sim M_\pi^2 /(4\pi F_\pi )^2 \sim 1/100$. Since a virtual photon can
never leave a diagram, the local contact terms only start at dimension two.
One has four and 17 terms at second and third order, respectively. Many of these
are simple renormalization of masses and coupling constants and can be absorbed
accordingly. At leading order, the virtual photons modify the covariant
derivative and the axial operator $u_\mu$, but these contributions can only appear
in loop graphs. For the construction of this effective Lagrangian and a more
detailed discussion of the various terms, see ref.\cite{ms}

Let me just present the pertinent results of that paper here. First, Weinberg's
finding concerning the scattering length difference $a(\pi^0 p) - a(\pi^0 n)$,
which is entirely given by a dimension two term $\sim m_u -m_d$, could be confirmed.
This is not surprising because for the case of neutral pions there is no contribution
from the em Lagrangian of order two and three. To third order there is no term,
because the charge matrix has to appear quadratically and never two additional pions
can appear. This will change at fourth order. Second, it was noted that the 
I--spin breaking terms in $a(\pi^0 p)$ can be as large as the I--spin conserving ones,
\be
a(\pi^0 p) = a(\pi^0 p)_{\rm str, IC} +  a(\pi^0 p)_{\rm str, IV} + 
a(\pi^0 p)_{\rm em} = (-0.48 - 0.11 - 0.29) \,\,\, ,
\ee
for the values of the LECs as determined in ref.\cite{bkmnpa} It is also important
to note that the em effects are entirely due to the pion mass difference, cf.
Fig.3, given by the operator $C\langle QUQU^\dagger\rangle$.
\begin{center}
\begin{figure}[hbt]
\centerline{
\psfig{figure=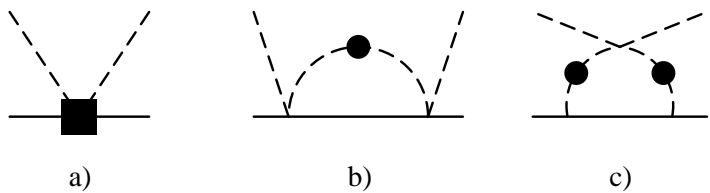,height=1.0in}
}
\caption{Graphs contributing to isospin violation in  $\pi^0$--proton 
   scattering. Solid and dashed lines denote nucleons and pions, in
   order. The heavy dot and the box refer to the 
   em counterterm at order $e^2$, i.e. the term proportional to $C$,
   and the dimension two strong insertion $\sim m_u -m_d$, respectively. 
   Diagram a) has previously been considered by 
   Weinberg. \label{fig:a0p}}
\end{figure}
\end{center}
Another quantity of interest is the pion--nucleon $\sigma$--term. Of course,
one now has to differentiate between the one for the proton and the one for
the neutron, whose values to third order differ by the strong neutron--proton
mass difference,
\be
\sigma_n (0) = \sigma_p (0) + 4B_0 (m_u - m_d) c_5 = \sigma_p (0) 
+ (m_n - m_p)_{\rm str}~,
\ee
with $B_0 = |\langle 0 |\bar{q}q|0\rangle | /F_\pi^2$ proportional to the scalar quark
condensate and $c_5 < 0$ a dimension
two LEC (for more details, see e.g. \cite{bkmnpa,ms}). For the proton and the 
same LECs used before, one finds
\be
\sigma_p(0) = \sigma_p^{IC} (0) + \sigma_p^{IV} (0) 
= 47.2~{\rm MeV} - 3.9~{\rm MeV}  = 43.3~{\rm MeV} \,\,\, ,
\ee
which means that the isospin--violating terms reduce the proton
$\sigma$--term by $\sim 8\%$. The electromagnetic effects are again
dominating the isospin violation since the strong contribution 
is just half of the strong
proton--neutron mass difference, 1~MeV. Furthermore, one gets
$\sigma_p (2M_{\pi^+}^2) - \sigma_p (0) = 7.5$~MeV, which 
differs from the result in the isospin limit (7.9~MeV) 
by 5\% and is by about a factor
two too small when compared to the dispersive analysis of ref.~\cite{gls}
This small difference of 0.4~MeV is well within the uncertainties related to the
so--called remainder at the Cheng--Dashen point.\cite{bkmcd}
Again, these corrections should be considered indicative since a fourth
order calculation is called for. Furthermore, the channels involving
charged pions need to be investigated for the reasons discussed above.
It is worth to emphasize again that we have a consistent machinery at hand to
simultaneously calculate the strong and the em isospin violating effects.

\section{Outlook}

I have presented first steps towards a systematic analysis of electromagnetic
and strong isospin breaking effects in the pion--nucleon system at low
energies. To third order in small momenta, a consistent machinery exists
and first results have been obtained. In ref.\cite{ulfosa} isospin violation
in the context of charged and neutral pion photoproduction was discussed.
Specifically, to third and fourth order the calculations by 
Bernard et al.\cite{bkmph} include the pion mass difference and it was argued
that this is the dominant isospin breaking effect. Indeed, to third order
using the effective Lagrangian obtained in ref.\cite{ms}, it can be shown for
$\gamma N \to \pi^0 N$ that the only em isospin violation comes from the 
rescattering graph with an insertion $\sim C$ on the internal pion line
(cf. fig.3b with the incoming $\pi^0$ line substituted by a photon), which is
nothing but the pion mass difference.\cite{gm} This means that the important 
cusp effect due to the opening of the $\pi^+ n$ threshold appears already at third order,
lending credit to the calculations of ref.\cite{bkmph} Nevertheless, fourth order
terms are important to understand the absolute magnitude of the electric dipole
amplitude $E_{0+} ( \gamma N \to \pi^0 N )$, whereas the energy dependence of this
quantity is only mildly affected by higher order terms.

Clearly, only by combining the precise machinery with the accurate data
from pionic atoms and pion photoproduction we can hope to get another
bound on the quark mass difference $m_u - m_d$. In a more ambitious step
to follow one should consider the near threshold data for $\pi N \to \pi \pi N$,
which encode (among other things) information on the elusive $\pi\pi$ scattering
lengths( see e.g.\cite{bkmpp}). 
Unfortunately, now that so much progress has been made on the
theoretical side, most of the meson factories are closing down. It remains
to be seen whether or not the presently available data for the various
processes mentioned can be understood to sufficient accuracy and whether
these data themselves are accurate enough to allow to pin down such fine
effects like strong isospin violation.

\section*{Acknowledgments}
I thank the organizers for  providing a stimulating atmosphere. I am
very grateful to V\'eronique Bernard, Nadia Fettes, 
Norbert Kaiser, Guido M\"uller and Sven Steininger
for pleasant collaborations.

\end{document}